
\documentstyle{amsppt}
\magnification=\magstep1
\nologo
\NoBlackBoxes
\define\Diff{\operatorname{Diff}}

\define\QS{\operatorname{QS}}
\define\QC{\operatorname{QC}}
\define\Homeo{\operatorname{Homeo}}
\topmatter
\title
UNIVERSAL TEICHM\"ULLER SPACE\\ IN GEOMETRY AND PHYSICS
\endtitle
\author
Osmo Pekonen
\endauthor
\affil
Department of Mathematics \\ University of Jyv\"askyl\"a \\ P.O. Box
35 \\ SF-40351 Jyv\"askyl\"a \\ Finland
\endaffil
\endtopmatter
\document

\bigskip

{\bf Abstract.} {\it Lipman Bers' universal Teichm\"uller space,
classically  denoted by $T(1)$, plays a significant role in
Teichm\"uller theory, because all the Teichm\"uller spaces $T(G)$
of Fuchsian groups $G$ can be embedded into it as complex
submanifolds. Recently, $T(1)$ has also become an object of
intensive study in physics, because it is a promising geometric
environment for a non-perturbative version of bosonic string
theory. We provide a non-technical survey of what is currently
known about the geometry of $T(1)$ and what is conjectured about its
physical meaning. Our bibliography should be rather
comprehensive, but we apologize for any unjustified omissions.}

\bigskip
\noindent{\bf 1. Some classes of homeomorphisms}
\bigskip

Let $\widehat{\Bbb C}={\Bbb C} \cup \lbrace \infty \rbrace$  be
the extended complex plane. We shall denote the unit disc $\{ z
\in \widehat{\Bbb C}:\;| z| < 1 \}$ by $\Delta$, the unit sphere
$\{ z \in \widehat{\Bbb C}:\;\mid z\mid = 1 \}$ by $S^1$, and the
exterior of the unit disc $ \{ z \in \widehat{\Bbb C}:\;| z| > 1
\} = \widehat{\Bbb C} \setminus (\Delta \cup S^1) $ by $
\Delta^{\star}. $

A homeomorphism $w: D\rightarrow w(D)$ between domains in
$\widehat{\Bbb C}$ is {\it quasiconformal} (qc) [3, 50] if and
only if $w$ has locally integrable generalized derivatives
satisfying almost everywhere on $D$ the {\it Beltrami equation}
$$
w_{\bar z}(z)=\mu(z)\,w_{z}(z) \tag 1.1
$$
for some measurable complex function $\mu$ on $D$ called the {\it
Beltrami differential}  with
$$
\underset {z\in D}\to {\operatorname{ess\,sup}}\, | \mu
(z)|=\parallel \mu\parallel_{\infty}<1.\tag 1.2
$$
A solution of (1.1) is called $\mu$-{\it conformal}\,; in  the
special case $\parallel \mu \parallel_{\infty} = 0$, $w$ is {\it
conformal}, i.e., biholomorphic.

Geometrically, quasiconformality means that $f$ maps any
infinitesimally small circle to an ellipse whose ratio of the
major axis to the minor axis is uniformly bounded by some number
$K<\infty$ called  the {\it maximal dilatation}. Such an $f$ is
called $K$-{\it quasiconformal.} The relationship between the
number $K$ and the function $\mu$, also called the {\it complex
dilatation} of $f$, is as follows:
$$
K=\sup_{z\in D}\,\frac{1+|\mu (z)|}{1-|\mu (z)|}.\tag 1.3
$$

Denote the space of Beltrami differentials  $ L^{\infty}(D)_{1}; $
it is the open unit ball in the complex Banach space
$L^{\infty}(D)$ of essentially bounded functions in $D$. The
existence and uniqueness, up to three prescribed values, of
solutions for the equation (1.1) with an arbitrary Beltrami
differential is guaranteed by a fundamental theorem due to Gauss,
Morrey, Bojarski [14], and Ahlfors and Bers [4]. In this survey,
we omit most proofs, but many of the deeper aspects of the theory
depend on careful analysis of the solutions of (1.1).

Let us think of the upper half-plane as a hemisphere of the
Riemann sphere $\widehat{\Bbb C}={\Bbb P}^{1}{\Bbb C}$. An
increasing self-homeomorphism $f$ of  the real axis $\Bbb R$   is
called {\it quasisymmetric} (qs) if it can be extended to a {\it
quasiconformal} mapping of the upper half-plane $\Bbb H$ that
fixes the point at infinity. Beurling and Ahlfors [13] showed
that $f$ is quasisymmetric, if for some constant $K$, $1\leq
K<\infty$,
$$
\frac 1K \leq \frac{f(x+t)-f(x)}{f(x)-f(x-t)}\leq K\tag 1.4
$$
for all real $x$ and positive $t$. More precisely, such an $f$ is
called  $K$-{\it quasisymmetric} (K-qs).

We may always switch from $\Bbb H$ to $\Delta$ via the  Cayley
transform $z\mapsto\frac{z-i}{z+i}$ which maps $(0,1,\infty)$  to
$(-1,-i,1)$, respectively. The explicit identification of $\Bbb
R$ to $S^1$ is given by
$$
x = -\cot \frac{\theta}{2},\quad \text{ or},\quad
e^{i \theta} = \frac{x-i}{x+i}. \tag 1.5
$$
A continuous vector field $u(e^{i \theta}) \frac{\partial}
{\partial \theta}$ on $S^1$ becomes, on the real line, $F(x)
\frac{\partial}{\partial x}$ with
$$
F(x) = \frac{1}{2} (x^2 + 1)\, u(\frac{x-i}{x+i}). \tag  1.6
$$
Conversely,
$$
u(e^{i \theta}) = \frac{2F(x)}{x^2+1}=2\,{\sin}^{2}
\,\frac{\theta}{2}\,F(-\cot \frac{\theta}{2}). \tag 1.7
$$
In particular, if $u$ vanishes at $(-1,-i,1)$, we see that
$$
F(0) = F(1) = 0\quad  \text{and}\quad \frac{F(x)}{x^2+1}
\rightarrow 0\;\text{as}\;x \rightarrow \infty~. \tag 1.8
$$

In analogy with the half-space model, an  orientation-preserving
self-homeomorphism $f$ of the unit circle $S^{1}$ is called {\it
quasisymmetric} (qs) if it can be extended to a quasiconformal
(qc) mapping of the unit disc $\Delta$. In the disc model, the
$K$-qs condition is most conveniently given in terms of the cross
ratio
$$
(z_{1},z_{2},z_{3},z_{4})=\frac{z_{4}-z_{1}}{z_{4}-
z_{2}}\, \frac{z_{3}-z_{2}}{z_{3}-z_{1}}.\tag 1.9
$$
We need to require that
$$
\frac{1}{2K}\leq \left(f(z_{1}),f(z_{2}),f(z_{3}),f(z_{4})\right)
\leq 1-\frac{1}{2K}\tag 1.10
$$
whenever $(z_{1},z_{2},z_{3},z_{4})=1/2$. The explicit constant
$K$ may not be the same in (1.3), (1.4) and (1.10), but this is
irrelevant for our present purposes. Again, a homeomorphism $f$
is qs if it is $K$-qs for some $K$.  In the sequel, we can
equivalently deal with quasisymmetric maps on $S^{1}$ or $\Bbb
R$, whichever seems to be more convenient, and we sometimes use
the symbol $X$ to designate either of these spaces. We shall then
denote the group of qs maps of the space $X$ by $\QS (X)$.

In the disc model, the M\"obius group $\text{M\"ob} (S^{1})$
consists of the boundary transformations induced by  the
conformal automorphisms of $\Delta$, the M\"obius transformations
$A: \Delta\rightarrow\Delta$,
$$
Az=\lambda\,\frac{z-a}{1-\bar{a} z}\tag 1.11
$$
with $z\in \Delta $, $|\lambda|=1$ and $| a| <1$. The M\"obius
group is a 3-dimensional subgroup of $\QS (S^{1})$  isomorphic to
$\text{PSU}(1,1;{\Bbb C})$.

Moreover, $\text{M\"ob} (S^{1})$ acts on the left on both $\QS (S^{1} )$
and $\QC (\Delta )$, the space of quasi\-conformal self-mappings
of $\Delta$. To see this, just notice that the cross ratio (1.9)
is M\"obius-invariant.

In the half-space model, the M\"obius group is $\text{PSL}(2;{\Bbb
R})$, the group of maps of the form
$$
f(z)=\frac{az+b}{cz+d}\tag 1.12
$$
with real coefficients $a,b,c,d$ satisfying $ad-bc=1$.

M\"obius trasformations can also be described as the solutions of
the differential equation
$$
Sf=0,\tag 1.13
$$
where $S$ is the {\it Schwarzian derivative},
$$
Sf=\left( \frac{f''}{f'}\right)' -\frac{1}{2} \left(
\frac{f''}{f'}\right)^{2}.\tag 1.14
$$
Direct computation gives the transformation rule
$$
S(f\circ g)=(Sf\circ g)(g')^{2}+Sg,\tag 1.15
$$
so that precomposition by a M\"obius transformation $f$ leaves the
Schwarzian derivative of a smooth function $g$ invariant. The
Schwarzian derivative will later show up in various contexts.

The extension of a qs map to a qc map is by no means unique. The
extension operator constructed by Beurling and Ahlfors [14], who
worked in the half-space model, has the drawback that it is not
{\it conformally natural}. Working in the disc model, Tukia [80]
and later also Douady and Earle [27] defined a conformally
natural extension operator $E: \QS (S^1 ) \rightarrow \QC (\Delta
)$ which satisfies the required naturality condition  $E( A\circ
f)= A\circ E(f)$ for any $A\in \text{M\"ob} (S^1)$ and $f\in \QS
(S^1 )$. A simpler construction with refined estimates for the
maximal dilatation has been given by Partyka [63]. A conformally
natural extension operator is not unique either: another one
$\tilde E$ is readily obtained by putting $\tilde E f = (E(f^{-
1}))^{-1}$.

In particular, if $f$ is a diffeomorphism on $S^{1}$, then it
surely can be continued as a diffeomorphism, a fortiori, as
a qc map, to the closed unit disc. Hence, smooth implies qs. We
shall denote the group of $C^{\infty}$  orientation-preserving
diffeomorphisms of $X$ by $\Diff (X)$.

The following chain of subgroup inclusions summarizes the most
important classes of homeomorphisms
$$
\text{M\"ob} (X) < \Diff (X) < \QS (X) < \operatorname{Homeo} (X).\tag 1.16
$$

Various other interesting spaces of homeomorphisms (real-analytic,
H\"older, symmetric [34],...) could be designated. For
our present purposes, let us introduce, following Zygmund [87],
the $\Lambda^{*}$ class
$$
\align \Lambda^{*} ({\Bbb R}) = \{ & F
:  {\Bbb R} \rightarrow {\Bbb R}\,|\, F \;\text{is continuous,
satisfying normalizations (1.8);} \; \text{and},  \\ & | F(x+t) +
F(x-t) -2F(x) | \leq C | t | \;\text{for some constant}\; C,  \\
& \text{for all}\; x\;\text{and}\; t\; \text{real}. \} \endalign
$$
Then $\Lambda^{*}({\Bbb R})$ is a non-separable Banach space
under the {\it Zygmund norm} which equals, by definition, the
best constant $C$ for $F$. Namely,
$$
|| F || = {\sup_{x,t}} \left | {{F(x+t) + F(x-t) - 2F(x)} \over
{t}} \right | \tag 1.17
$$

The interest of the {\it Zygmund class} $\Lambda^{*}({\Bbb R})$
lies in the fact that, according to Reimann [72], the Zygmund
class  comprises precisely the vector fields for quasisymmetric
flows on ${{\Bbb R}}$.

\bigskip

\noindent{\bf 2. Geometric quantization of bosonic string theory}

\bigskip

We shall now discuss some physics as a motivation for further
mathematical developments. {\it Bosonic string theory} [26,
39, 60] is a proposal of unified field theory where the
elementary particles called bosons are supposed to appear as
1-dimensional extended objects in the Planck scale; hence,
topologically they look like either $\Bbb R$ (open string) or
$S^{1}$ (closed string). We shall work with closed strings.  The
string hypothesis introduces a new symmetry group into physics,
the group $\Homeo (S^{1})$, as this is the internal symmetry
group of a closed string. {\it Non-perturbative bosonic string
theory}  would be  based, ideally at least, on the group $\Homeo
(S^{1})$. We would like to geometrize this group, but as it seems
to be intractable, in practice, we need to content ourselves with
some subgroup.

There is a standard procedure in physics called {\it geometric
quantization} [84] to pass from a classical system to a quantum
system. In the classical system, the {\it observables} are
functions  $f$ in the {\it phase space} which is a smooth
manifold $M^{2n}$ endowed with a symplectic form $\omega$ ;  in the
corresponding quantum system the observables need to be converted
into operators $T_{f}$ acting in some Hilbert space in such a way
that Poisson brackets of functions are converted into Lie
brackets of operators
$$
T_{\lbrace f_{1},f_{2}\rbrace }=\lbrack T_{f_{1}},
T_{f_{2}}\rbrack.\tag 2.1
$$
The standard way to achieve this is to produce a Hermitian line
bundle $\Cal L$ over $M$ with a Hermitian connection  $\nabla$
whose curvature equals $\omega$. $\Cal L$ exists if and only if
$\omega$ represents an integral cohomology class. Then the sought-for
operators will be given by
$$
T_{f}= -i\nabla_{X_{f}}+f\tag 2.2
$$
where $X_{f}$ is the Hamiltonian vector field corresponding to
the observable $f$  by the formula $X_{f}=-\omega^{-1}(df,.)$.
The operators $T_{f}$ act in the Hilbert space of
square-integrable sections of $\Cal L$ with respect to the
canonical volume form $\frac{\omega^{n}}{n!}$ of $(M,\omega)$. In
fact, up to this point, we have only achieved {\it
prequantization} while the difficult  {\it Dirac problem}
concerning the irreducibility of the representation $f\mapsto
T_{f}$ remains to be settled. This final step in the geometric
quantization programme can often be achieved by introducing a
K\"ahler structure on the phase space and restricting to the
holomorphic square-integrable sections.

Geometric quantization of string theory involves many unsolved
problems which we shall discuss in due course later on. In any
case, to get started we could try to produce a symplectic
structure on $\Homeo (S^{1})$ or, more modestly, on $\Diff
(S^{1})$. The Lie algebra of the  infinite-dimensional Lie group
$\Diff (S^{1})$ is the algebra $\operatorname{Vect}(S^1 )$ of
smooth vector fields on  the circle. These have Fourier modes
labeled by the ring of integers  $\Bbb Z$, so that
$\operatorname{Vect}(S^1 )$ formally behaves like an
{\it odd}-dimensional space; hence, certainly $\Homeo (S^1 )$ or
$\Diff (S^{1})$ as such cannot carry a symplectic structure.
Heuristically, an odd number of degrees of freedom need to be
removed before we can expect a symplectic phase space.

The simplest idea is to remove the Fourier zero mode by
quotienting  away the group of rotations
$\operatorname{Rot}(S^{1})$, or the circle itself. The resulting
moduli space is denoted by
$$
N=\Diff (S^{1}) / \operatorname{Rot}(S^{1}).\tag 2.3
$$
Bowick and Rajeev [17, 19-22] discovered that the space $N$
carries, indeed, the structure of an infinite-dimensional
K\"ahler manifold.  We shall explicit this K\"ahler structure
later on, but implicitly, this phenomenon may be understood as an
infinite-dimensional analogue of the  finite-dimensional standard
argument of Kirillov, Kostant, and Souriau [46] which produces a
symplectic structure in the coadjoint orbit space of a Lie group
acting on the dual of its Lie algebra.

Moreover, there exists another obvious "even-dimensional"
quotient space, namely
$$
M=\Diff (S^{1}) / \text{M\"ob}(S^{1}).\tag 2.4
$$
Then $N$ is a holomorphic disc bundle over $M$. The
Kirillov-Kostant-Souriau argument applies to $M$ as well, and,
indeed, Bakas [9] and Witten [83] have proved that the dual of
the Lie algebra of $\Diff (X)$ admits no other coadjoint actions
by non-trivial subgroups of $\Diff (X)$. Hence, in principle, we
can choose either $N$ or $M$ as the underlying phase space of our
geometric quantization scheme. We shall see that $M$ is far more
interesting.

We shall not review the method of coadjoint orbits, but let us
mention  that the dual of $\operatorname{Vect}(S^1)$ can be
identified with the space of {\it Hill operators} $H_{u}$ acting
on smooth functions on the circle,
$$
H_{u}=\left( \frac{d}{dz}\right)^{2}+u(z),\tag 2.5
$$
where $u$ is any smooth function on the circle [9]. Then, under
arbitrary smooth reparametrizations $z\rightarrow f(z)$, the Hill
operators transform as follows:
$$
H_{u}\rightarrow \frac{1}{f'^{2}}\left( \frac{d}{dz}\right)^{2}-
\frac{f''}{f'^{3}}\,\frac{d}{dz}+u(f(z)).\tag 2.6
$$
Explicit calculation shows that (2.6) is equivalent to
$$
H_{u}\rightarrow M(f'^{-\frac{3}{2}})\,H_{\tilde u}\, M(f'^{-
\frac{1}{2}}),\tag 2.7
$$
where $M(.)$ is the multiplication operator and $H_{\tilde u}=
\left( \frac{d}{dz}\right)^{2}+\tilde u(z)$ is given by
$$
H_{\tilde u} (z)=\left( \frac{d}{dz}\right)^{2}+
f'^{2}u(f(z))+\frac{1}{2}\,Sf(z),\tag 2.8
$$
where $S$ is the Schwarzian derivative (Segal [74]).

\bigskip

\noindent{\bf 3. Universal Teichm\"uller space $T(1)$}

\bigskip

Our fundamental sequence of inclusions (1.16) can be quotiented into
$$
M=\Diff (X)/\text{M\"ob} (X) \subset QS (X)/\text{M\"ob}
(X)\subset \operatorname{Homeo} (X)/\text{M\"ob} (X).\tag 3.1
$$
Here we recognize a classical object; namely,
$$
T(1):=QS (X)/\text{M\"ob} (X)\tag 3.2
$$
is the {\it universal Teichm\"uller space} that Bers introduced
in [10, 11]. Equivalently, we may think of $T(1)$ as the space of
quasisymmetric homeomorphisms of the circle, say, which have
three prescribed values. We stipulate the three points $\pm 1$
and $-i$ to be fixed. The infinite-dimensional space  $T(1)$ is
{\it universal} in the sense that it  contains as subspaces all
the other Teichm\"uller spaces whose definition we briefly recall
next [13, 43, 49, 54, 78].

Let $G$ be a {\it Fuchsian group}, i.e., a discrete subgroup of
$\text{M\"ob} (X)$. The {\it Teichm\"uller space} $T(G)$ is
defined by
$$
T(G)=\lbrace [f]\in T(1)\,|\, f\circ \gamma \circ f^{-1}\in
\text{M\"ob} (X)\; \text{for all}\; \gamma \in G \rbrace.\tag 3.3
$$
These spaces are partially ordered: $G<G'$ clearly implies
$T(G')\subset T(G)$; in particular, all the Teichm\"uller spaces
$T(G)$ are contained in the universal one. (The "1" in the
notation $T(1)$ refers  to the trivial group.) Moreover, the
inclusion $T(G) \subset T(1)$ turns out to be a holomorphic embedding.

Another approach to Teichm\"uller theory is global analytic.
The {\it Teichm\"uller space}
$T(\Sigma)$ of a Riemann surface $\Sigma$ is defined as the
parametrization space of its complex structures up to isotopy. A
complex structure may be given as a smooth section $J$ of the
endomorphism bundle of the tangent bundle of the surface $\Sigma$
which is an anti-involution, i.e.,
$$
J^2 =-\operatorname{id}.\tag 3.4
$$
Denote the space of all complex structures $J$ by $\Cal A$. Of
course, $\operatorname{Diff}(\Sigma)$ acts on $\Cal A$ via the
pull-back operation. In the case of a compact orientable surface
of genus $>1$, the Teichm\"uller space $T(\Sigma )$ simply equals
the moduli space ${\Cal A}/{\operatorname{Diff}_{0}(\Sigma)}$
where $\operatorname{Diff}_{0}(\Sigma)$ is the identity component
of $\operatorname{Diff}(\Sigma)$. This simple global analytic definition is
powerfully exploited in the treatise [78].

In the above set-up, we can easily define a K\"ahler structure on
$T(\Sigma)={\Cal A}/{\operatorname{Diff}_{0} (\Sigma)}$. By
differentiating the relation (3.4), we see that the tangent
vectors $\dot J$ of $T(\Sigma)$ at $J$ anticommute with $J$ under
composition of endomorphisms. Hence, the formula
$$
\omega ({\dot J}_{1},{\dot J}_{2})= \int_{\Sigma}\,
\operatorname{tr}\,(J\circ {\dot J_{1}}\circ {\dot J_{2}})\tag 3.5
$$
defines a 2-form on $\Cal A$. The integration is with respect to
the hyperbolic metric uniquely corresponding to $J$. This
correspondence is natural with respect to the action of
$\operatorname{Diff}(\Sigma)$; a fortiori, with respect to the
action of the subgroup $\operatorname{Diff}_{0}(\Sigma)$, so it
passes to the quotient. Moreover, it is straightforward to check
that the resulting 2-form on the Teichm\"uller space $T(\Sigma)$
is non-degenerate and closed; hence, a K\"ahler form. This
K\"ahler form (up to a constant multiple) is the classical
Weil-Petersson K\"ahler 2-form.

If the surface $\Sigma$ is uniformized by a Fuchsian group $G$,
then the two definitions coincide:
$$
\Sigma = X/G\,\,\implies\,\,T(\Sigma)=T(G).\tag 3.6
$$
This is a theorem of Tukia [80] which Douady and Earle [27]
reproved using their conformally natural extension operator.

We shall briefly indicate how $T(\Sigma)$ and $T(G)$ are related
to each other. For this purpose, we need to discuss two pertinent
classes of solutions for the Beltrami equation (1.1)
corresponding to specially chosen Beltrami differentials $\mu \in
L^{\infty} (\widehat {\Bbb C})_{1}$.

{\bf The real-analytic $w_{\mu}$-theory:}  By
applying the fundamental existence and uniqueness theorem to the
Beltrami differential which is $\mu$ on $\Delta$ and is extended
to $\Delta^{\star}$ by reflection $(\tilde{\mu} (1/\bar{z}) =
\overline{\mu(z)} z^{2}/\bar{z}^{2}$ for $z \in \Delta)$, one
obtains the quasiconformal homeomorphism $w_{\mu}$ of $\Bbb C$
which is $\mu$-conformal in $\Delta$, fixes $\pm 1$ and $-i$, and
keeps $\Delta$ and $\Delta^{\star}$ both invariant.

{\bf The complex-analytic $w^{\mu}$-theory:}
By applying the existence and uniqueness theorem to the Beltrami
differential which is $\mu$ on $\Delta$ and zero on
$\Delta^{\star}$, one obtains the quasiconformal homeomorphism
$w^{\mu}$  on $\Bbb C$, fixing $0,1,\infty$, which is
$\mu$-conformal on $\Delta$ and conformal on $\Delta^{\star}$.

It is a fact that $w_{\mu}$ depends only real-analytically on $\mu$,
whereas $w^{\mu}$ depends complex-analytically on $\mu$. The latter
extension is so useful that it carries the name "Bers' trick".

Let us now try to make (3.6) look a little more plausible. First
of all, a Beltrami differential $\mu$ is $G$-{\it equivariant}, if
it is compatible with the action of $G$ on $\Delta$; more
precisely, this leads to the requirement
$$
\mu (\gamma z) {\bar{\gamma}}'(z)/\gamma'(z)=\mu (z)\tag  3.7
$$
which should hold almost everywhere on $\Delta$ for every $\gamma
\in G$. Let us denote the space of $G$-compatible Beltrami
differentials $L^{\infty}(G)$. An alternative description of
$T(G)$ can now be given as
$$
T(G)= L^{\infty}(G)/\sim \tag 3.8
$$
where $\mu\sim\nu$ if and only if $w_{\mu}=w_{\nu}$ on
$\partial\Delta=S^{1}$, which happens if and only if $w^{\mu}=w^{\nu}$
on $\Delta^{*}\cup S^{1}$.

Now, if $\mu$ is $G$-invariant, then $w_{\mu}$ conjugates
$G$ to another Fuchsian group
$$
G_{\mu}=w_{\mu}Gw_{\mu}^{-1}.\tag 3.9
$$
The equivalence class of $\mu$ in $T(G)$ represents the Riemann
surface $X_{\mu}=\Delta/ G_{\mu}$.

In the reverse direction, one can use $w^{\mu}$ to conjugate $G$
to a {\it quasi-Fuchsian group}
$$
G^{\mu}=w^{\mu}G(w^{\mu})^{-1}\tag 3.10
$$
so that $G^{\mu}$ acts discontinuously on the {\it quasidisc}
$\Delta^{\mu}=w^{\mu}(\Delta)$ and its exterior
$\Delta^{*\mu}=w^{\mu}(\Delta^{*})$. Now, the Riemann surface
$X_{\mu}$ is represented by $\Delta^{\mu}/G^{\mu}$ (whereas
$\Delta^{*\mu}/G^{\mu}$ is the fixed Riemann surface
$\Delta^{*}/G$, since $w^{\mu}$ is conformal on $\Delta^{*}$).

\bigskip

\noindent{\bf 4. Models of $T(1)$}

\bigskip

We can think about $T(1)$ in several ways. The following three
classical models of $T(1)$ are the best-known:
\roster
\item"{\bf (a)}" {\it the real-analytic model} consisting of all
M\"obius-normalized quasisymmetric homeomorphisms of the unit
circle $S^1$;
\item"{\bf (a')}" {\it the geometric model} consisting of all
M\"obius-normalized {\it quasicircles}, i.e., all images of the
standard circle under a global quasiconformal map that fixes the
points $\pm 1$ and $-i$;
\item"{\bf (b)}" {\it the complex-analytic model} comprising all
functions which fix $0,1,\infty$, which are  univalent
on the exterior of the unit disc $\Delta^{*}$ and which allow
quasiconformal extension to the whole Riemann sphere.
\endroster

For yet other models, see [45, 47].

Specializing (3.8), we may also define the universal Teichm\"uller
space as a quotient of Beltrami differentials:
$$
T(1) = {L}^{\infty} (\Delta)_{1} / \sim  \tag 4.1
$$
where $\mu \sim \nu$ if and only if $w_{\mu} = w_{\mu}$ on
$\partial \Delta = S^{1}$, or equivalently, if and only if
$w^{\mu}$ and $w^{\nu}$ coincide on $\Delta^{\star} \cup S^{1}$.

We let
$$
\Phi : L^{\infty}(\Delta)_{1} \longrightarrow T(1) \tag 4.2
$$
denote the quotient projection.  $T(1)$ inherits its canonical
structure as a complex Banach manifold from the complex structure
of $L^\infty(\Delta)_{1}$; indeed, $\Phi$ becomes a holomorphic
submersion.

The derivative of $\Phi$ at $\mu = 0$ :
$$
d_{0} \Phi : L^\infty(\Delta) \longrightarrow T_{0} T(1) \tag 4.3
$$
is a complex-linear surjection whose kernel is the space $N$ of
``infinitesimally trivial Beltrami differentials''
$$
N = \{ \mu \in L^{\infty}(\Delta) : \int_{\Delta} \mu \phi = 0~~
\text{ for all}~~~\phi \in A(\Delta) \} \tag 4.4
$$
where $A(\Delta)$ is the Banach space of $L^1$ integrable
holomorphic functions on the disc.  Thus, the tangent space at
the origin $0= \Phi(0)$ of $T(1)$ is $L^{\infty}(\Delta)/N$.

It is now clear that to $\mu \in L^{\infty}(\Delta)_{1}$ we
can associate the quasisymmetric homeomorphism
$$
f_{\mu} = w_{\mu} \mid_{S^{1}} \tag 4.5
$$
as representing the Teichm\" uller point $[\mu]$ in the
real-analytic model (a) of $T(1)$.  Indeed $T(1)_{(a)}$ is the
homogeneous space :
$$
\align T(1)_{(a)} = & \QS (S^{1}) / \text{M\"ob} (S^1)  \\  = &
\{ \text{quasisymmetric homeomorphisms of}\; S^1\;
\text{fixing}\; \pm 1\; \text{and}\; -i \} \endalign
$$

In the geometric model (a') of $T(1)$, we  think of the points of
$T(1)$ as the images of $S^{1}$ under $w^{\mu}$.

There is a plethora of characterizations of the quasidiscs and
their boundaries, the quasicircles [37]. Perhaps the most elegant
is Ahlfors' condition [2] which identifies quasicircles among
those Jordan curves of the complex plane which pass through
$\infty$ (this can be achieved by a M\"obius transformation):
Such a Jordan curve $C$ is a quasicircle if and only if there is
a constant $M$ such that for any three distinct points $a,b,c$ on
$C$ with $b$ between $a$ and $c$
$$
|b-a| \leq M |c-a|.\tag 4.6
$$
A generic quasidisc turns out to be a fractal object.

Alternatively, $[\mu]$ is represented by the univalent function
$$
f^\mu = w^\mu \mid_{\Delta^{\star}} \tag 4.7
$$
on $\Delta^{\star}$, in the complex-analytic model (b) of $T(1)$.
A more natural choice of the univalent function representing
$[\mu]$ is to use a different normalization for the solution
$w^{\mu}$ (since we have the freedom to post-compose by a
M\"obius transformation).  In fact, let
$$
W^{\mu} = M^\mu \circ w^\mu \tag 4.8
$$
where $M^\mu$ is the unique M\"obius transformation so that the
univalent function (representing $[\mu]$) :
$$
F^\mu = W^\mu \mid_{\Delta^{\star}} \tag 4.9
$$
has the properties:
\roster
\item"{(i)}"  $F^\mu$ has a simple pole of residue 1 at $\infty$ ;
\item"{(ii)}" $(F^{\mu}(z) - z) \rightarrow 0$ as $z \rightarrow \infty$.
\endroster
Thus, the expansion of $F^\mu$ in $\Delta^{\star}$ is of the form:
$$
F^\mu(z) = z + {{b_{1}} \over {z}} + {{b_{2}} \over {z^{2}}} +
{{b_{3}} \over {z^{3}}}+\ldots \tag 4.10
$$
Let us note that the original ($0,1,\infty$ fixing) normalization
gives an expansion of the form:
$$
f^\mu(z) = z (a + {{\beta_{1}} \over {z}} + {{\beta_{2}} \over
{z^{2}}} + {{\beta_{3}} \over {z^{3}}} + \ldots) \tag 4.11
$$
and the M\" obius transformation $M^\mu$ must be $M^{\mu }(w) =
w/a -\beta_{1}/a$.  Since $(a,\beta_{1},\beta_{2},..)$  depend
holomorphically on $\mu$, we see that $(b_{1},b_{2},b_{3},...)$
also depend  holomorphically on $\mu$.  Thus, our complex-analytic
version of $T(1)$ is:
$$
\align T(1)_{(b)} = \{ & \text{Univalent functions in} \;
\Delta^{\star}\;  \text{with power series of the form}\; (4.10),
\\  & \text{allowing quasiconformal extension to the whole plane}
\}. \endalign
$$

In the general theory of univalent functions, the functions of the
type (4.10) are known as the class $\Sigma\,$ [28]. It is not difficult
to compute that the area of the corresponding quasidisc is
$$
A=\pi\, (1-\sum _{n=1}^{\infty}\, n\,| b_{n}| ^{2}). \tag 4.12
$$
Of course, this is non-negative so that we deduce the classical
Area Theorem [28, 49, 54] about the coefficients $b_{n}$ in the
class $\Sigma$:
$$
\sum _{n=1}^{\infty} n\,| b_{n}| ^{2}\leq 1.\tag 4.13
$$

We may think of the coefficients $b_{n}$ as coordinates on
$T(1)$. A  refinement of the Area Theorem shows that $b_{n}={\Cal
O}(n^{-c})$ with $c=0.509...$ [23], but the coefficients in the
class $\Sigma$ still retain many mysteries. It is known that $|
b_{1}|\leq 1\,$, $| b_{2}|\leq\frac 23$, and $|b_{3}|\leq\frac{1}{2}
+ e^{-6}.$ These bounds are {\it sharp} but there is no
$\lq\lq$Bieberbach conjecture" about the general sharp upper
bound for $| b_{n}|.$ Nonetheless, we can think of $T(1)$ as a
space of certain sequences $(b_{1}, b_{2}, b_{3},\dots )$.

\bigskip

\noindent{\bf 5. The physicist's wish-list}

\bigskip

In this chapter, we explain why the universal property of $T(1)$
makes it  an attractive object of study from the point of view of
{\it non-perturbative} bosonic string theory whose precise geometric
formulation, as  we should stress, is unknown. First we shall review
in simple non-technical terms the basic ideas of the prevailing
{\it perturbative} bosonic string theory to the benefit of the
reader who is not familiar with the physics literature.

The central issue in any quantum field theory is to evaluate the
{\it partition function} $Z$ which gives the quantum-mechanical
probability amplitudes of the system under study. Feynman
introduced in 1948 a quantization scheme where $Z$ is computed as
a {\it path integral} over the space of paths representing the
possible worldlines of elementary particles. The possible
spacetime trajectories of a propagating  pointlike elementary
particle are 1-dimensional paths,  whereas a propagating bosonic
{\it string}, or a  1-dimensional extended object, sweeps out
2-dimensional  world-surfaces.  A natural generalization of the
Feynman path integral then is an integral over all possible
world-sheets. As a first approximation, one can limit to deal with
compact orientable  surfaces which are topologically classified
by the genus $\gamma =0,1,2,\dots$ The emergence of a handle in
the propagation pattern corresponds to the breaking apart of two
strings; correspondingly, the annihilation of two strings closes
a handle. In reality,  one should also take into account the
degenerate situations where, e.g., a handle shrinks to a node.

While the topological classification of compact orientable
surfaces is easily understood, their geometrical diversity is
more intricate. The possible geometries are given as the
infinite-dimensional cone $\Cal M$ of all  Riemannian metrics $g$
of the underlying topological surface. This space, however, is
physically redundant. The physically meaningful space in each
genus $\gamma$  is the parametrization space of conformal
structures, or, the {\it Riemann moduli space} ${\Cal M}_{\gamma}$.
It is well-known that ${\Cal M}_{0}$ is a point, ${\Cal M}_{1}=
{\Bbb H}/\text{PSL}(2;{\Bbb Z})$, while ${\Cal M}_{\gamma}$ is an
orbifold of dimension $6\gamma -6$. The Riemann moduli
space of a surface  $\Sigma$ admits as its covering space the
{\it Teichm\"uller space} $T(\Sigma)$, the parametrization space
of conformal structures up to isotopy. Conformal structures are
the relevant {\it intrinsic} geometries of the surface $\Sigma$,
so one should develop an integration scheme over the moduli. Some
explicit results are known, e.g., the volumes of some
Teichm\"uller spaces in the Weil-Petersson metric (Penner [66]).

We think of the string propagating in a fixed background space
that, as a first approximation, can be taken to be the flat
Minkowski space of some dimension $D$.  We also need to take into
account the {\it extrinsic} geometries of the string; in other
words, the various ways in which a string may be embedded into
the ambient space-time. An extrinsic metric is induced on the
surface $\Sigma$ as a pull-back of the flat background metric via an
embedding $s$.  Thus we should also integrate over all embeddings
$s:\Sigma\rightarrow {\Bbb R}^{D-1,1}$.

Consequently, we should be looking for the partition function in
the form of a perturbative series
$$
Z=\sum_{\gamma =0}^{\infty}\,\int_{{\Cal M}_{\gamma}}\,e^{-S}\,.\tag 5.1
$$
Here $S=S(g,s)$ is the {\it Polyakov energy}, i.e., the Dirichlet
energy of an arbitrary embedding of the propagating string into
the background spacetime. The integration in (5.1) is with
respect to the so-called {\it Polyakov measure} over the moduli
space ${\Cal M}_{\gamma}$ of each genus $\gamma$ and also over
the infinite-dimensional space of all embeddings $s$.  Polyakov
discovered in 1981 that this measure  exhibits {\it conformal
anomaly cancellation} in the critical spacetime dimension $D=26$.

Perturbative bosonic string theory suffers from several drawbacks.
First of all, the summation over the genus in (5.1) is well-known
to be divergent [40, 66]. Secondly, the need to prescribe the
topology and geometry of the background spacetime is philosophically
unsatisfactory. The spacetime should rather arise as an
excitation. So far, merely the critical dimension $D=26$ arises
as a constraint. Thirdly, the critical dimension is outlandish,
be it lowered to the slightly more palatable $D=10$ in {\it
superstring theory} [26, 39] which incorporates fermions as well.

Perhaps these drawbacks indicate that we are just scratching the
surface of some underlying intrinsic geometric principle that
would imply more stringent conditions on the global properties of
spacetime. The proper geometric environment of bosonic string
theory should be some kind of $\lq\lq$universal Riemann moduli
space" which would comprise the moduli of surfaces with an
arbitrary number of handles, cusps, boundary components, and
nodes. One heuristic candidate for such an object has been put
forward by Friedan and Shenker [32, 33], but it is not
mathematically well-established.

The only classically known universal moduli space in mathematics
literature is Bers' universal Teichm\"uller space $T(1)$,
although no viable notion of universal Riemann moduli space
corresponds to it. The potential physical interpretation of
$T(1)$ as a superspace, in the sense of DeWitt and Wheeler, was
discussed by Bers already in [12]. From the modern point of view,
$T(1)$ is bound to be a highly relevant object, as it plays a
role in both of the existing approaches to the quantization of
bosonic strings:
\roster
\item"{(i)}" From the point of view of {\it perturbative} bosonic
string theory, $T(1)$ contains as subspaces all the finite-dimensional
Teichm\"uller spaces corresponding to various perturbative orders.
\item"{(ii)}" From the point of view of {\it non-perturbative} bosonic
string theory, $T(1)$ is contained in the space $\Homeo
(S^1)/\text{M\"ob} (S^1)$ which, in principle, should be the
ultimate arena of the geometric quantization of bosonic string
theory.
\endroster

Perhaps the perturbative series (5.1) ought to be replaced by a
single integral over the universal Teichm\"uller space
$$
Z= \int_{T(1)}e^{-S}.\tag 5.2
$$
Some preliminary  speculations about how the measure in $T(1)$
should look like in terms of the coefficients $b_{n}$ appear in
[42], but no actual progress has been recorded.

The physicists' wish-list for mathematicians to achieve the
geometric quantization of $T(1)$ includes (at least) the
following items [67]:
\roster
\item
{\it Universal geometry:} $T(1)$ should be a K\"ahler manifold
whose K\"ahler form $\omega$ pulls back to the Weil-Petersson
form on each classical Teichm\"uller space $T(G)$.
\item
{\it Universal topology:} There should be an action by a
$\lq\lq$universal mapping class group" on $T(1)$ which pulls back
to each $T(G)$.
\item
{\it Universal line bundle:} There should exist a Hermitian line
bundle $\Cal L$ over $T(1)$ with a connection whose curvature
equals $\omega$.  \item  {\it Universal measure:} $T(1)$ should
carry a $\lq\lq$Haar  measure" with respect to which the
classical locus, or the union of the images of the embeddings of
all the $T(G)$, is dense and of measure zero in $T(1)$.
\item
{\it Universal action principle:} There should exist a scalar-valued
function $S$ ($\lq\lq$universal Polyakov energy") whose
gradient flow should have a superset of the classical locus as
attracting fixed points. This function should coincide with a
K\"ahler potential of the universal Weil-Petersson K\"ahler form.
\endroster

We shall review the above-listed items emphasizing the
established aspects of the theory. The current state of art seems
to be that the item (1) is well-established for the quotient
space $\Diff (S^{1})/ \text{M\"ob} (S^{1})$ while some
evidence of its validity has been advanced in a suitable
subspace of $T(1)$ (Nag and Sullivan [58]) and, from a different
point of view, even in $\Homeo (S^{1}) / \text{M\"ob} (S^{1})$ (Penner
[67]).

The solution to the item (2) has been claimed by Penner [67], but
the discussion of his graph theoretic methods would bring us too
far. Ratiu and Todorov [71] suggested that Quillen's determinant
line bundle construction [70] applied to a family of Cauchy-Riemann
operators parametrized by $M$ could solve the item (3). However,
one sees with difficulty how Quillen's construction could be extended
to $T(1)$. Possibly, ordinary calculus ought to be replaced by
$\lq\lq$quantum calculus" in the sense of Alain Connes' non-commutative
geometry [24].

The item (4) has been preliminarily discussed by Wiesbrock [81, 82]
and by Nag and Sullivan [58]. Some time ago, we suggested in [65]
that the quasidisc area functional $A$ on $T(1)$ might serve as a
heuristic candidate for the universal action principle required
in item (5), because the quadratic expression in  (4.12) can be
interpreted as the Dirichlet energy  of the harmonic extension in
$\Delta$ defined by the boundary values of the univalent function
$F^{\mu}$ in (4.10). However, we have not been able to compare the
functional $A$ with the Polyakov energy of each genus.

We have not yet pointed out the existence of a natural distance
function on $T(1)$. Denote by $K(h)$ the minimal dilatation of a
quasiconformal self-mapping of $\Delta$ with the same boundary
values as $h$. Then the {\it Teichm\"uller metric} on $T(1)$ is
defined by
$$
d_{1}(f,g)=\frac{1}{2}\log\, K(f\circ g^{-1}),\quad [f], [g] \in
T(1).  \tag 5.3
$$
Obviously, the value $d_{1}(f,g)$ does not change if we replace
$f$ by $A\circ f$ and $g$ by $B\circ g$, where $A$ and $B$ are
conformal mappings, and hence the Teichm\"uller metric is
well-defined. Minimizing the dilatation only over $G$-compatible
quasiconformal mappings, one analogously obtains a Teichm\"uller
metric $d_{G}$ for each $T(G)$. On the other hand, the metric
space  $(T(1), d_{1})$ induces a metric space structure to each
subspace $T(G)$. The Teichm\"uller metric $d_{1}$ on $T(1)$ is
universal in the sense that it induces the same topology as
$d_{G}$ to each $T(G)$. Moreover, $d_{1}\leq d_{G}$ and,
according to Strebel [77], in general, $d_{1}< d_{G}$. The
Weil-Petersson metric is non-complete, while the Teichm\"uller
metric is complete. In particular, the two metrics are not
equivalent. Indeed, there is no Riemannian metric corresponding
to the Teichm\"uller metric, so that it does not provide an
answer to the item (1) in the physicists' wish-list above.

The introduction of $\Diff S^1$ in string theory was originally
motivated as a globalization of the work of Frenkel, Garland, and
Zuckerman [31] who gave the conditions for the consistency of
string theory in terms of a certain Lie algebra cohomology of
vector fields of the circle. The {\it algebraic} approach to
string theory is a vast topic which is beyond the scope of this
survey. Let us mention, though, that the need to understand better
also the algebraic relationship between Polyakov's perturbative
approach and the non-perturbative geometric quantization approach
has been emphatically expressed by Manin [51].

\bigskip

\noindent{\bf 6. The tangent space of $T(1)$}

\bigskip

In order to do differential geometry on $T(1)$, we first need to
describe its tangent space in the various models.

{\bf Tangent space to the real-analytic model}: Since
$T(1)$ is a homogeneous space (according to the model (a)) for
which the right translation by any fixed quasisymmetric
homeomorphism acts as a biholomorphic automorphism, it is enough
in all that follows to restrict attention to the tangent space at
a single point of $T(1)$, the origin, or, the class of the
identity homeomorphism.

Given any $\mu \in L^\infty (\Delta)$, the tangent vector
$d_{0}\Phi(\mu)$ is represented by the real vector field $V[\mu]
= \dot{w}[\mu] {{\partial} \over {\partial z}}$ on the circle
that produces the 1-parameter flow $w_{t \mu}$ of quasisymmetric
homeomorphisms:
$$
w_{t \mu}(z) = z + t \dot{w} [\mu](z) + o(t) \tag 6.1
$$
The vector field becomes in the $\theta$-coordinate :
$$
V[\mu] = \dot{w}[\mu](z) {{\partial} \over {\partial z}} =
u(e^{i \theta}) {{\partial} \over {\partial \theta}} , \tag 6.2
$$
where,
$$
u(e^{i \theta}) = {{\dot{w}[\mu](e^{i \theta})} \over {i e^{i \theta}}}~.
\tag 6.3
$$
By our normalization, $u$ vanishes at $\pm 1$ and $-i$.

As mentioned before, the Zygmund class $\Lambda^{*} ({\Bbb R})$
comprises precisely the vector fields for quasisymmetric  flows
on $\Bbb R$.  Hence, the tangent space to the real-analytic model
of $T(1)$ becomes :
$$
\align T_{0} (T(1)_{(a)}) = \{ u(e^{i \theta})
\frac{\partial}{\partial \theta}\, : (i)\; &  u : S^1 \rightarrow
{\Bbb R}\;\text{is continuous,  vanishing at}\; (\pm 1,-i) ; \\
(ii)\; & F_{u}(x) = \frac{1}{2}\, (x^2 +1)\, u{(\frac{x-i}{x+i}
)}\quad \text{is in}\;\;\Lambda^{*}({\Bbb R})\;  \} \tag 6.4
\endalign
$$

We will say that a continuous function $u: S^1 \rightarrow {\Bbb
R}$ is in {\it{the Zygmund class}} $\Lambda^{*}(S^1)$ {\it{on the
circle}}, if, after adding the requisite $(ce^{i \theta} +
\bar{c}e^{i \theta} + b)$ to normalize $u$, the function
satisfies the conditions in (6.4).

{\bf Tangent space to the complex-analytic model}:  A
tangent vector at $0$ (the identity mapping) to $T(1)_{(b)}$
corresponds to a 1-parameter family $F_{t}$ of univalent
functions (each allowing quasiconformal extension):
$$
F_{t}(z) = z + {{b_{1}(t)} \over {z}} +  {{b_{2}(t)} \over
{z^{2}}} +  {{b_{3}(t)} \over {z^{3}}} +
\ldots,~~\text{in}~~\vert z \vert > 1~, \tag 6.5
$$
with
$$
b_{k}(t) = t\dot{b}_{k}(0) + o(t),\quad k = 1,2,3, \ldots\tag 6.6
$$
The sequences $\{ \dot{b}_{k}(0)~,~k \geq 1 \}$ arising this
way uniquely correspond to the tangent vectors.

Applying Ahlfors' deep infinitesimal theory for solutions of the
Beltrami equation [1], Nag [57] was able to characterize which
sequences occur in (6.5). To announce his result, let us expand
as a Fourier series the vector field $V[\mu]$ :
$$
u(e^{i \theta}) = {{\dot{w}[\mu](e^{i \theta})} \over {ie^{i \theta}}} =
\sum_{k=-\infty}^{\infty} a_{k} e^{ik \theta}. \tag 6.7
$$
Since $u$ is real valued, one knows $a_{-k} = \bar{a}_{k},
k \geq 1$. The coefficients $a_{o}$ and $a_{\pm 1}$ do not matter
owing to the $sl(2,{{\Bbb R}})$ normalization.

Nag [57] established the following interesting identitities between
the coefficients $a_{k}$ in (6.7) and $\dot{b}_{k}$ in (6.6):
$$
\dot{b}_{k}(0) = i a_{-k} = i \bar{a}_{k}, \quad
\text{for every} \quad k \geq 2 . \tag 6.8
$$
This immediately implies a precise description of the tangent
space to $T(1)$ in the complex-analytic model:  In (6.5)
precisely those sequences  $\left( \dot{b}_{1}(0),
\dot{b}_{2}(0), \dot{b}_{3}(0), \ldots \right)$ occur for which
the function
$$
u(e^{i \theta}) = i \sum_{k=1}^{\infty} \bar{\dot {b}}_{k}(0)
e^{ik\theta} -i \sum_{k=1}^{\infty} {\dot {b}}_{k}(0) e^{-
ik\theta} \tag 6.9
$$
is in the Zygmund class on $S^1$.

\bigskip

\noindent{\bf 7. The almost complex structure of $T(1)$}

\bigskip

The Lie algebra of the Lie group $\Diff (S^{1})$ is the Lie
algebra $\operatorname{Vect}(S^1 )$  of smooth vector fields on
$S^{1}$. The complexification  $\operatorname{Vect}_{\Bbb C}(S^1
)$  of $\operatorname{Vect}(S^1 )$  is generated by the Fourier
modes
$$
L_{n}=e^{in\theta}\frac{d}{d\theta}=iz^{n+1}\frac{d} {d
z},\,\quad n\in {\Bbb Z} \tag 7.1
$$
with $z=e^{i\theta}$. To $\operatorname{Vect}_{\Bbb C}(S^1 )$
there does not correspond any global Lie group, yet, Neretin [61]
has constructed a complex semigroup whose tangent cone is a
convex cone in $\operatorname{Vect}_{\Bbb C}(S^1 )$. The Lie
bracket of $\operatorname{Vect}_{\Bbb C}(S^1 )$ is given by the
Witt law
$$
[L_{m},L_{n}]=i(n-m)L_{m+n}.\tag 7.2
$$
A tangent vector to the orbit space $M=\Diff (S^{1})/\text{M\"ob}
(S^1)$ at its origin is a linear combination
$$
\vartheta =\sum_{m\neq 0,\pm 1}\vartheta_{m}L_{m},\,\,\,\bar
\vartheta_{m}=\vartheta_{-m},\tag 7.3
$$
where $\vartheta=u(\theta)\frac{\partial}{\partial\theta}$ is the
corresponding smooth real vector field on the circle and the
$\vartheta_{m}$ are the Fourier coefficients of $u(\theta)$.  The
Lie algebra corresponding to the three missing modes
$\vartheta_{-1}, \vartheta_{0},\vartheta_{1}$, is {\it
sl}$(2;{\Bbb R})$, of course. We may conjugate the series (7.3)
by the conjugation operator $J$ to
$$
J\vartheta=\sum_{m\neq 0,\pm 1}-
i\operatorname{sgn}(m)\vartheta_{m}L_{m}.\tag 7.4
$$
This is again a smooth vector field, but $J$ can be applied to a
much wider class: A classical result of Zygmund [88] says that
conjugation of Fourier series preserves the Zygmund class
$\Lambda^{*}(S^1)$.

Notice that $J^{2}=-\text{id}$. Kerckhoff (unpublished) first
pointed out the fact  that the conjugation operation on Zygmund
class vector fields on $S^1$ transmutes to the almost complex
structure of $T(1)$.  Nag [57] applied the identities (6.8) to
give a simple proof of this fact. Indeed, we need to prove that
the vector field $V[\mu]$ in (6.2) is related to $V[i \mu]$ as a
pair of conjugate Fourier series. But the tangent vector
represented by $\mu$ in the complex-analytic  description of
$T(1)$ corresponds to a sequence $(\dot{b}_{1}(0),
\dot{b}_{2}(0), \dot{b}_{3}(0), \ldots )$, as explained above.
Since the $b_{k}$ are holomorphic in $\mu$, the tangent vector
represented by $i \mu$ corresponds to $(i \dot{b}_{1}(0), i
\dot{b}_{2}(0), i \dot{b}_{3}(0), \ldots)$.  The relation (6.8)
immediately shows that the $k^{\text{th}}$ Fourier coefficient of
$V[i \mu]$ is $-i.sgn(k)$ times the $k^{\text{th}}$ Fourier
coefficient of $V[\mu]$, as required.

\bigskip

\noindent{\bf 8. The Bers embedding of $T(1)$}

\bigskip

To provide a system of complex coordinates for $T(1)$ Bers [11]
embedded $T(1)$ as a holomorphically convex domain into the
complex Banach space $B$ which consists of all functions $\phi
(z)$, holomorphic in the lower half-plane, $\Bbb L$, with bounded
{\it Nehari norm} defined by
$$
\parallel\phi\parallel =\underset {z\in {\Bbb L}} \to
{\operatorname{ess\,sup}} \,4\,\vert y^{2}\phi(z)\vert.\tag 8.1
$$
In the complex-analytic model of $T(1)$,  let $f^{\mu}$ represent
a point of $T(1)$. We think of $f^{\mu}$ as a quasiconformal map
which is univalent in $\Bbb L$. The Bers
embedding  $T(1)\hookrightarrow B$ is defined by
$$
f^{\mu}(z) \mapsto S(f^{\mu})(z), \quad z\in {\Bbb L}\tag 8.2
$$
where $S$ is the Schwarzian derivative as in (1.14)
which annihilates M\"obius moves according to (1.13).

Since an element $f^{\mu}$ is determined by
the Beltrami differential $\mu$ up to a M\"obius move, we may
think of the Bers embedding as a function of $\mu$ as well. It
then defines a holomorphic embedding of $T(1)$ into $B$ with
respect to the complex structure of the Beltrami differentials.

It is an interesting problem to study the locus of $T(1)$ in $B$
in the Bers embedding.  In particular, we may compare the locus
of $T(1)$ to the bigger locus $S$ in $B$ of the Schwarzian
derivatives of {\it all} univalent maps on $\Bbb L$. The
following facts are known:

(i) In the Nehari norm, $T(1)$ contains an open ball of radius 2
and is contained in a closed ball of radius 6 ([49, 54]).

(ii) $T(1)$ is an open set in $B$ while $S$ is closed in $B$, and
the closure of $T(1)$ is a proper subset of $S$ (Gehring [35]).

(iii) The interior of $S$ is $T(1)$ (Gehring [36]).

(iv) $T(1)$ is connected (Earle-Eells [29]) while $S$ contains
isolated points (Astala [5], Astala-Gehring [6]).

(v) Suppose that $h$ is a univalent map of $\Bbb L$ onto a simply
connected domain $D$ of hyperbolic type in $\Bbb C$.  Then $S(f)$
is in the closure of $T(1)$ if and only if for each $K>1$ there
exists a homeomorphism $g$ of $D$ onto a quasidisc such that for
each disc $Q$ in $D$, $g\vert Q$ has a  $K$-quasiconformal
extension to $\widehat {\Bbb C}$ (Astala and  Gehring [7]).

(vi) $T(1)$ is contractible [29] but not star-shaped (Krushkal
[48]).

Tukia [79] has showed how to embed $T(1)$ as a real analytic
convex domain in a real Banach space.

\bigskip

\noindent{\bf 9. The K\"ahler structure of $T(1)$}

\bigskip

Nag and Verjovsky [59] proved that the natural inclusion
$$
M=\operatorname{Diff}(S^{1})/\text{M\"ob}(S^{1})\hookrightarrow T(1)\tag 9.1
$$
is holomorphic. The proof amounts to showing that, if we write in
(7.4)
$$
J\vartheta = u^{*}(\theta)\frac{\partial}{\partial \theta},\tag 9.2
$$
then $u^{*}$ is essentially the Hilbert transform of $u$; this is
not very difficult.

A more subtle result of Nag [59] endows a subspace of $T(1)$ with
a K\"ahler structure and shows the inclusion (9.1) to be a
K\"ahler isometry onto its image. Recall that the existence of a
symplectic form $\omega$ on $M$ is predicted by the theory of
coadjoint orbits. To compute it explicitly, we impose the
condition $d\omega= 0$, or, equivalently, at the origin
$$
\omega ([L_{m},L_{n}], L_{p})+\omega ([L_{n},L_{p}], L_{m})+
\omega ([L_{p},L_{m}], L_{n})=0.\tag 9.3
$$
Also, $\omega$ must vanish whenever one of its arguments is
$L_{0}, L_{\pm 1}$ since these vector fields give the zero
tangent vector to $M$. The conditions (7.2) and (9.3) now lead to
a system of difference equations whose only possible solution
readily yields a homogenous K\"ahler form $\omega$ which is given
at the origin by
$$
\omega (L_{m}, L_{n})=\alpha (m^{3}-m)\delta_{m,-n},\,\,\,m,n\in
{\Bbb Z}\setminus \lbrace 0, \pm 1\rbrace.\tag 9.4
$$
The constant $\alpha\in {\Bbb C}\setminus \lbrace 0\rbrace$ is
arbitrary.

Let $v=\sum_{m}v_{m}L_{m}$ and $w=\sum_{m}w_{m}L_{m}$ of the form
(7.3) represent two tangent  vectors to $M$ at the origin.
Then the K\"ahler metric $g$, whose K\"ahler form $\omega$ was
determined above, assigns the inner product
$$
g(v,w)=-2i\alpha \operatorname{Re}\Big(\sum_{m=2}^{\infty}\,v_{m}
{\overline w}_{m} (m^{3}-m)\Big).\tag 9.5
$$
According to standard results in harmonic analysis,  the Fourier
coefficients of a $C^{k+\epsilon}$ smooth function on $S^1$ decay
at least as fast as $1/n^{k+\epsilon}$. Hence, the infinite
series in (9.5) converges absolutely whenever the vector fields
$v$ and $w$ are $C^{3/2+\epsilon}$ smooth on $S^1$ for any
$\epsilon >0$.  Zygmund class functions are not necessarily
smooth at all, so that the series (9.5) does not yield a
well-defined inner product on all of $T(1)$. Claims have been made,
though, that even $\Homeo (S^{1})/ \text{M\"ob} (S^1 )$ carries a
K\"ahler structure in some sense [67].

The K\"ahler structure $\omega$ is universal in the sense that it
is closely related to the Weil-Petersson K\"ahler forms on each
$T(G)$. However, the relationship is not by simple restriction of
domains from the  infinite-dimensional space $T(1)$ to the
complex-analytic subspace $T(G)$, because $T(G)$ is {\it
transversal} to the leaf $M$ of the foliation of $T(1)$ in the
following sense: Let us use the geometric definition of $T(1)$ as
the space of M\"obius-normalized quasidiscs. Bowen [16] proved
the deep result that if $G$ uniformizes a compact Riemann surface,
then every non-origin point of $T(G)$ corresponds to a quasidisc
with {\it fractal} boundary. On the other hand, the quasidiscs
corresponding to points of $M$ are the ones with $C^{\infty}$
boundaries (Kirillov [45]). Nag [59] showed that every non-null
tangent vector to $T(G)$ at the origin produces a vector field on
$S^1$ that cannot be even $C^{3/2+\epsilon}$ smooth.

Nonetheless, the expression of the metric (9.5) is formally the
same as that of the Weil-Petersson metric even when it diverges,
and it can be regulated as explained by Nag in [59]. This procedure
is not entirely satisfactory from the point of the physicists'
wish-list, however.

\bigskip

\noindent{\bf 10. Curvature properties of
$\text{Diff}(S^{1})/\text{M\"ob} (S^{1})$}

\bigskip

Conformal field theory [44] suggests that the natural value of
the constant $\alpha$ is  $\alpha =\frac{1}{12}$; see Atiyah [8]
for a purely topological derivation. This normalization is
natural also when $\omega$ is viewed as the generator of the
second Gelfand-Fuks cohomology $H^{2}(\operatorname{Vect}^{\infty}
(S^{1});{\Bbb C})$ (Segal [74]). Besides being the unique
symplectic form on $M$, the 2-cocycle $\omega$ can also be found
as the unique central extension of the Lie algebra of vector
fields on the circle. The centrally extended Lie algebra then is
called the {\it Virasoro algebra}. It is customary to write
$\alpha = \frac{c}{12}$ so that the Virasoro law reads as
$$
[L_{m},L_{n}]=i(n-m)L_{m+n}+\frac{c}{12}(m^3-m)\delta_{m,-n}.\tag
10.1
$$
Remarkably, the spectrum of the values of the coupling constant
$c$ admitting unitary representations of the Virasoro algebra
(10.1) is continuous precisely for $c>1$  [38].  For $c<1$, the
spectrum consists of the discrete series
$$
c=1-\frac{6}{(k+2)(k+3)},\quad k=1,2,3,\dots\tag 10.2
$$
This phenomenon also shows that the value $c=1$, i.e.,  $\alpha =
\frac{1}{12}$ is critical.

The Ricci curvature of the K\"ahler manifold $M$ has been
computed by Bowick and Lahiri [18].  The method of Toeplitz
operators for dealing with infinite-dimensional Ricci curvature
was introduced by Freed [30]. Such computations are surprising
because an infinite-dimensional trace can be performed without
any regularization. For the critical normalization
$\alpha=\frac{1}{12}$ one obtains
$$
\operatorname{Ricci}=-26\times \omega .\tag 10.3
$$
It is surprising to see the mysterious critical dimension $D=26$
of bosonic string theory emerge in this context! The critical
occurrence of the number $26$ in two seemingly disparate roles
must be  an instance of the subtle interplay of Feynman's
quantization and geometric quantization of bosonic string theory
rather than a mere numerical coincidence, yet this phenomenon has
never been geometrically explained.

The K\"ahler structures of the orbit space $N=\Diff (S^{1})/ S^1$
form a two-parameter family:
$$
\omega(L_m, L_n)=(am^{3} +bm)\,\delta_{m,-n}.\tag 10.4
$$
This is  non-degenerate when either $a=0, b\neq 0$, or $a\neq 0,
-b/a\neq n^2$ with $n\in {\Bbb Z}$. For $a=0$, the infinite-dimensional
trace in the Ricci curvature of $(N,\omega)$ diverges, while for
$a\neq 0$, it is finite and the result is
$$
R_{\bar{m} n }=(-\frac{26}{12}
m^{3}+\frac{1}{6}m)\delta_{m,n}.\tag 10.5
$$
In fact, this computation was made earlier than that for $M$ by
several authors [17, 19-22, 64, 86]. Mickelsson [52, 53] extended
the formula (10.5) to the case of a string moving on a simple
compact Lie group.

The formula (10.5) has been extended to the supersymmetric set-up
as well [41, 62, 68, 73, 85]. Then the critical dimension $D=10$
of superstring theory arises in an equally mysterious manner. The
notion of universal super-Teichm\"uller space seems not to have
been developed, though.

\bigskip

\noindent{\bf 11. Holomorphic embedding of $T(1)$ in the
universal Siegel disc}

\bigskip

Consider the Sobolev space $\Cal H= H^{1/2} (S^1 , {\Bbb
R})/{\Bbb R}$ of all $H^{1/2}$ real functions on the circle
modulo the constant maps and its complexification ${\Cal H}_{\Bbb
C}= H^{1/2} (S^1 , {\Bbb C})/{\Bbb C}$.  Harmonic analysis tells
us that the Fourier series
$$
f(e^{i\theta})=\sum_{n=-\infty}^{\infty} u_{n}e^{i n\theta} \tag
11.1
$$  of a $H^{1/2}$ function $f$ converges quasi-everywhere,
i.e., off some set of capacity zero. We may think of ${\Cal
H}_{\Bbb C}$ equivalently as the space $\ell_{2}^{1/2}$ of
complex sequences $(\dots ,u_{-3}, u_{-2}, u_{-1}, u_{0}=0,
u_{1}, u_{2}, u_{3}, \dots )$ such that $\{ \sqrt{\vert n\vert}
u_{n} \}$ is square summable.
The Hilbert transform on $T(1)$ given by (7.4) also
extends to ${\Cal H}_{\Bbb C}$.

The fundamental orthogonal decomposition of ${\Cal H}_{\Bbb C}$
is given by
$$
{\Cal H}_{\Bbb C}=W_{+}\oplus W_{-} \tag 11.2
$$
where $W_{+}$ (resp\. $W_{-}$) consists of those functions $f\in
{\Cal H}_{\Bbb C}$ whose negative (resp\. positive) index Fourier
coefficients vanish.

We may provide $\Cal H$ with a symplectic structure, i.e., a
non-degenerate  skew-symmetric bilinear form $S$ whose formula is
$$
S(f, g)=\frac{1}{2\pi}\int_{S^1}f(e^{i\theta}) \frac{d
}{d\theta}g(e^{i\theta})\,d\theta.\tag 11.3
$$
The same formula extends to ${\Cal H}_{\Bbb C}$ as well. The
$W_{+}$, resp\. $W_{-}$, are the $-i$, resp\. $+i$, eigenspaces
of the Hilbert transform. Let $f_{\pm}$ denote the
projection of $f$ to $W_{\pm}$. The inner product on ${\Cal
H}_{\Bbb C}$ is given by
$$
\langle f,g \rangle = iS(f_{+}, \bar {g}_{+}) -i S(f_{-}, \bar
{g}_{-}).\tag 11.4
$$

Then $\QS (S^1) $ acts faithfully on
$\Cal H$. The action $V$ of $\phi \in \QS (S^{1})$ on $f\in {\Cal
H}$ is given by
$$
V_{\phi} (f)=f\circ \phi -\frac{1}{2\pi}\int_{S^{1}}
f\circ \phi.\tag 11.5
$$
In fact, the class $\QS (S^{1})$ turns out to be the largest
possible class of homeomorphisms $\phi : S^{1} \rightarrow S^{1}$
for which $V_{\phi}$ maps $\Cal H$ to itself. Moreover, Nag and Sullivan [58]
show that the action $V$ preserves the canonical symplectic form $S$. On
the other hand, up to a constant multiple, $S$ turns out to be
the unique  $\text{M\"ob} (S^1)$-invariant, {\it a fortiori}, the
unique $\QS (S^{1})$-invariant symplectic form on $\Cal H$.

Hence, $\QS (S^1)$ becomes a subgroup of the group of real symplectic
automorphisms of the symplectic space $({\Cal H},S)$.
Moreover, $\operatorname{Sp}({\Cal H})/U(1)$ contains $T(1)=\QS
(S^1)/\text{M\"ob} (S^1)$ as an immersed subspace.

A {\it polarization} of the space $\Cal H$ with respect to $S$ is
a decomposition ${\Cal H}_{\Bbb C}=W\oplus \overline W$ such that
the complexification of $S$ takes zero values on arbitrary pairs
from $W$. The subspace $W$ is said to be {\it isotropic} for $S$. The
assignment
$$
\langle w_{1}, w_{2}\rangle = -iS({\bar w}_{1}, w_{2})\tag 11.6
$$
is a Hermitian inner product on $W$, and the decomposition is a
{\it positive polarization} if (11.6) is positive definite.  In this
case, $W$, its conjugate $\overline W$, and hence ${\Cal H}_{\Bbb
C}$ itself, can be completed to Hilbert spaces with respect to
the above Hermitian inner product. We may identify a positive
polarization with the isotropic subspace $W$ determining it. The
canonical positive polarization is given by $W_{+}$.

Note the fundamental fact that the image under the $\Bbb C$-linear
extension of a symplectic automorphism of a positive isotropic
subspace is again such a subspace. Hence,
$\operatorname{Sp}({\Cal H},S)$ acts transitively on the space of
all positive polarizations, and the stabilizer
subgroup at $W$ is evidently identifiable with the unitary group
$U(W,\langle.,.\rangle)$. It follows that the homogenous space
$\operatorname{Sp}/U$ can be identified with the family
$\operatorname{Pol}({\Cal H})$ of positive polarizations of $\Cal
H$. Either of these spaces can be easily identified with the
{\it universal Siegel disc}, denoted $S_{\infty}$:
$$
\align
S_{\infty}=\{ & \text{All bounded complex linear operators}\quad
Z: W_{+} \rightarrow W_{-} \\ & \text{such that}:\; (1)\; Z\;
\text{is symmetric w.r.t.}\; S,\; S(Zv,w)=S(Zw,v); \\ &
\text{and}\;(2)\; I-Z^{*}Z\; \text{is positive definite.} \}
\endalign
$$

The identification between $S_{\infty}$ and
$\operatorname{Pol}({\Cal H})$ is by associating to $Z\in
S_{\infty}$ the positive isotropic subspace $W$ which is the
graph of the operator $Z$. (Clearly, the origin in $D_{\infty}$
corresponds to the canonical polarization of $W_{+}$.) We have
seen that the universal Siegel disc can be described in the
following manners:
$$
S_{\infty}=\operatorname{Sp}/U=\operatorname{Pol}.\tag 11.7
$$
The positive polarizing subspace $W$ can also be taken to be the
$-i$-eigenspace of an arbitrary $S$-compatible
almost complex structure $J$ on $\Cal H$. $\lq\lq S$-compatible" means
that $J$ acts orthogonally with respect to $S$ and that the inner product
$\langle.,.\rangle = S(.,J(.))$ is positive definite. Thus, the
set of such $J$'s yields yet another description of $S_{\infty}$.
We stress that the symplectic structure $S$ on $\Cal H$ is completely
canonical while $J$ is not.

The {\it Grassmannian} $\operatorname{Gr}(W_{+}, {\Cal H}_{\Bbb
C})$ consists of all subspaces of ${\Cal H}_{\Bbb C}$ that are of
type $W_{+}$. Clearly, $S_{\infty}$ is embedded in
$\operatorname{Gr}$ as a complex subspace.  The symplectic form
$S$ extends to $\operatorname{Gr}(W_{+}, {\Cal H}_{\Bbb C})$. Nag
and Sullivan [58] (see also Nag [55, 56]) showed that the
following chain of mappings consists of equivariant holomorphic
symplectomorphisms:
$$
M\rightarrow T(1) \underset \Pi
\to{\rightarrow} S_{\infty}\rightarrow \operatorname{Gr}.\tag  11.8
$$

The problem of describing the image of $\Pi $ in $S_{\infty}$ is an
infinite-dimensional analogue of the classical Schottky problem.
Nag and Sullivan [58] provided a characterization of the locus
of $\Pi$  as the space of {\it multiplication-closed} polarizing
subspaces $W$ in ${\Cal H}_{\Bbb C}$. This means that for every
$f,g \in W$ such that the pointwise product function $fg$ minus
its mean value is in ${\Cal H}_{\Bbb C}$, that product is
actually in the given subspace $W$. This condition has a natural
interpretation in terms of Connes' non-commutative geometry
[24, 25, 58] but this would bring us too far. In view of  Shiota's
theorem [76], one expects a $\lq\lq$Novikov conjecture", possibly in
some non-commutative sense.  The locus of $T(1)$ in $S_{\infty}$
should be determined within $\operatorname{Gr}(W_{+}, {\Cal
H}_{\Bbb C})$  as those points $W\in \operatorname{Gr}(W_{+},
{\Cal H}_{\Bbb C})$ whose $\lq\lq$tau function" [69, 75]  satisfies
some special conditions related to the Korteweg - de Vries (KdV)
hierarchy of equations. That would tie in with the
finite-dimensional Novikov conjecture.

Such developments are beyond our scope, yet we briefly indicate
how the classical KdV equation for a smooth function $u=u(z,t)$
$$
\frac{\partial u}{\partial t}=\frac{\partial^{3}u}{\partial
z^{3}}+6u \frac{\partial u}{\partial z}\tag 11.9
$$
arises in the study of $M$. Indeed, the Lax form of (11.9) is
$$
\frac{\partial H_{u}}{\partial t}=3[ P_{u}, H_{u}], \tag 11.10
$$
where $H_{u}$ is the Hill operator $H_{u}=(\frac{d}{dz})^{2}
+u(z)$ as in (2.5) and
$$
P_{u}=\frac{4}{3}\partial^{3}_{z}+u\partial_{z}+\partial_{z}u.
\tag 11.11
$$
Thus, the KdV equation can be interpreted as describing flows
that  correspond to isospectral deformations of the Hill operator
[9, 74, 75].

\bigskip

\noindent{\bf 12. Conclusion}

\bigskip

There seems to be emerging a fascinating interplay of
Teichm\"uller theory and non-commutative geometry which may shed
new light on crucial issues of non-perturbative string theory. We
hope that our survey will be helpful for someone who is seeking
his or her way through the maze of existing literature before
tackling the forthcoming papers of Connes and Sullivan [25], Nag
and Sullivan [58], and Penner (in preparation).

\bigskip

{\bf Acknowledgement}

\bigskip

I have learned most of the material exposed here from
conversations and correspondence with Professor Subhashis Nag,
and from his papers and book. I am grateful for the hospitality
that I enjoyed during a stay with him at {\smc MATSCIENCE} in
Madras.

\enddocument